# Empirical Evaluations of Active Learning Strategies in Legal Document Review

Rishi Chhatwal, Esq.
Legal
AT&T Services, Inc.
Washington, D.C. USA
Email: rc134a@att.com

Nathaniel Huber-Fliflet
Legal Technology Solutions
Navigant Consulting, Inc.
Washington, D.C. USA
Email: nathaniel.huber-fliflet@navigant.com

Robert Keeling, Esq.
Complex Commercial Litigation
Sidley Austin LLP
Washington, D.C. USA
Email: rkeeling@sidley.com

Dr. Jianping Zhang
Legal Technology Solutions
Navigant Consulting, Inc.
Washington, D.C. USA
Email: jianping.zhang@navigant.com

Dr. Haozhen Zhao
Legal Technology Solutions
Navigant Consulting, Inc.
Washington, D.C. USA
Email: haozhen.zhao@navigant.com

*Abstract—* One type of machine learning, text classification, is now regularly applied in legal matters involving voluminous document populations because it can reduce the time and expense associated with the review of those documents. One form of machine learning – Active Learning – has drawn attention from the legal community because it offers the potential to make the machine learning process even more effective. Active Learning, applied to legal documents, is considered a new technology in the legal domain and is continuously applied to all documents in a legal matter until an insignificant number of relevant documents are left for review. This implementation is slightly different than traditional implementations of Active Learning where the process stops once achieving acceptable model performance. The purpose of this paper is twofold: (i) to question whether Active Learning actually is a superior learning methodology and (ii) to highlight the ways that Active Learning can be most effectively applied to real legal industry data. Unlike other studies, our experiments were performed against large data sets taken from recent, real-world legal matters covering a variety of areas. We conclude that, although these experiments show the Active Learning strategy popularly used in legal document review can quickly identify informative training documents, it becomes less effective over time. In particular, our findings suggest this most popular form of Active Learning in the legal arena, where the highest-scoring documents are selected as training examples, is in fact not the most efficient approach in most instances. Ultimately, a different Active Learning strategy may be best suited to initiate the predictive modeling process but not to continue through the entire document review.

## I. INTRODUCTION

Information management has become a significant business challenge, with the global volume of electronically stored information growing at a rapid pace (doubling roughly three times since 2010) [1]. In modern litigation, attorneys often face an overwhelming number of documents that must be reviewed and produced over the course of a matter. In large-scale litigation, legal teams may be required to produce millions of documents to opposing parties or regulators. These same teams typically must further wade through a sea of documents to find supporting evidence for their own arguments.

The costs involved in manually reviewing documents has grown dramatically as more and more information is stored electronically. This process requires extraordinary resources. Companies regularly spend millions of dollars producing responsive electronically stored documents for litigation matters [2]. Attorneys spend countless hours reviewing documents to respond to routine discovery requests, and these enormous costs are then passed along to clients. A RAND Corporation study revealed that the review process generates the bulk of discovery costs, with document review typically accounting for 73 percent of all production costs, collection consisting of about 8 percent of expenditures, and costs for data processing amounting to about 19 percent in typical cases [2].

To more efficiently cull massive volumes of data to identify responsive information, companies increasingly turn to machine learning techniques like text classification. In the legal domain text classification is typically referred to as predictive coding or technology assisted review (TAR). Predictive coding applies a supervised machine learning algorithm to build a predictive model that automatically classifies documents into predefined categories of interest. Attorneys typically employ predictive coding to identify so-called "responsive" documents, which are materials encompassed by document requests (or subpoenas) served upon the attorney's client by the counterparty to the litigation or investigation. In the course of identifying responsive documents, attorneys also typically review for materials that are covered by the attorney-client privilege or other types of privilege that typically protect documents from production. While less common, attorneys are increasingly employing machine learning to identify privileged documents in the data population.

Predictive coding, already highly valued in litigation settings, has been increasingly embraced in other legal matters such as responding to Department of Justice requests and mergers and acquisitions. As predictive coding's use increases, implementation methodologies have advanced. The earliest forms of predictive coding focused on lawyers making educated guesses and conducting random sampling to identify training documents used to teach the machine learning algorithm what it should look for. This approach is called Passive Learning, and while it can be effective, there is room for improvement. Recently, the legal community started using a more advanced technology to identify training documents and to enhance the efficiency of the overall predictive coding process called Active Learning in Machine Learning, also referred to as Continuous Active Learning (CAL) in the legal domain [3]. CAL is defined as an Active Learning process in which top-ranked training documents are selected for continual training and the process continues until there are not a significant number of responsive documents left for review. In this paper, we take a broad view of Active Learning, considering different training document selection strategies and if the process can be improved by stopping before the full CAL workflow completes. In the rest of the paper, the use of Active Learning refers to a broad interpretation of CAL that may use various training document selection strategies. Although there are many published articles and significant industry discussion about Active Learning in Machine Learning, there are few empirical studies about its effectiveness and implementation methodologies in legal document review, with some exceptions (e.g., [3][4][5]).

This paper reports on our empirical study of Active Learning in predictive coding. Our experiments, composed of five data sets from real legal matters across various industries, simulated real document review scenarios. Specifically, we studied two Active Learning training document selection approaches to identify which was the most effective at requiring the fewest number of documents to review and to determine if Active Learning is a superior learning methodology. Our work is unlike previously published studies on Active Learning in predictive coding (e.g., [3]). Until now, many in the legal community generally accepted that CAL (selecting top-ranked training documents and applying Active Learning until an insignificant number of relevant documents are left for review), is the best approach to develop predictive models. Our results suggest that Active Learning can effectively identify informative training documents quickly but that it becomes less effective over time depending on the document selection strategy. This indicates that Active Learning may be best suited to initiate the predictive modeling process but not to continue throughout the entire document review. In this paper, we (i) review Active Learning in the machine learning and legal domains; (ii) describe our experimental procedure and data sets; and (iii) report our results and findings, highlighting key components that have the largest influence on results.

## II. ACTIVE LEARNING

The earliest forms of predictive coding, as text classification is called in the legal domain, typically required attorneys to make a random selection to identify training documents that would teach a machine learning algorithm what to look for. While a random selection strategy is effective, it does not take advantage of attorney knowledge to target responsive content or the insight provided by the predictive model. This potentially may lead to inefficient model building and added document review burden on the attorney team – resulting in unneeded costs to the attorneys' clients.

Active Learning was proposed to make the identification of quality training exemplars much easier [6]. The purpose of Active Learning is to generate an accurate predictive model with as few training exemplars as possible. In Active Learning, the "learner" has control over choosing the training exemplars. Active Learning begins with a predictive model trained using a small set of labeled exemplars and then uses that model to select one or more informative training exemplars from a pool of unlabeled potential exemplars. The exemplars selected by the Active Learning algorithm are then reviewer-labeled (or coded) and added to the training set. A new model is then generated using the newly supplemented training set, which allows for the identification of new informative training exemplars. These steps are repeated until the results of the final model are acceptable or cannot be improved materially.

There are different Active Learning strategies for choosing training exemplars. The most popular strategy in the machine learning domain is to choose the most uncertain exemplars [7], these being the exemplars that the model finds most confusing. When creating linear models using machine learning algorithms such as Logistic Regression and Support Vector Machine with a linear kernel, uncertain exemplars are the exemplars that are close to the class decision boundary – meaning it's difficult for the learning algorithm to confidently place them in one of the two classes. Previous studies (e.g., [8]) used this strategy in their text classification tasks, and within the legal domain this strategy has been referred to as Simple Active Learning [5]. Another strategy called Top-Ranked Active Learning, and sometimes referred to as Continuous Active Learning (CAL) in the legal domain [3], chooses top-ranked, high-scoring exemplars. Choosing highly ranked exemplars has also been used to handle data sets with an imbalanced class distribution because there tend to be fewer uncertain positive exemplars (coding labels such as Responsive or Privileged) [9]. Clustering-based Active Learning is another strategy that targets training exemplars belonging to groups of various concept clusters [10].

In legal matters employing Active Learning, Top-Ranked Active Learning is the dominant strategy. It finds responsive documents quickly and does not require a validation set to test the performance of the model. This can reduce the time required to generate an effective model, in turn reducing attorney review costs and providing document content insights to more quickly develop a case strategy.

However, Top-Ranked still presents a research problem [3]. When an Active Learning selection strategy only selects the top-ranked (highest scoring) documents, it is possible that, at some point, these documents may not provide enough new information to improve the results of the model [11]. In such a situation, including lower-ranked documents could be useful to continue to improve the model's results because lower-ranked documents will contain additional diverse textual content.

While Top-Ranked can provide some advantages in document review, it also possesses disadvantages. Because Top-Ranked needs to repeatedly train models and score all documents, the project management and hardware resource costs can be high [3]. When dealing with tens of millions of documents, it is not practical to continue with a Top-Ranked Active Learning workflow until enough responsive documents are found. Imagine a matter with three million Responsive documents and each Top-Ranked round is 1,000 training documents; to complete a full Top-Ranked workflow, it would require at least 3,000 iterations of the process. To combat this challenge, Cormack and Grossman [5] have proposed a scalable version of the Top-Ranked strategy. Instead of using the entire document review population, their scalable version selects a large sample of the document review population and continues to generate models with new training documents until all the documents in the sample are exhausted. Implementing Top-Ranked with hundreds of reviewers can also prove challenging. It is imperative that the attorney review team agrees on the content that drives a Responsive decision. If there is disagreement or misunderstanding, the model may keep suggesting training documents that do not meet the desired content and move the review off course – a risk that is exacerbated as the size of the review team increases. Finally, Top-Ranked can add extra effort to the review process if a model stabilizes and cannot improve early in the Top-Ranked Active Learning workflow. Using the matter with three million Responsive documents as an example, implementing the Top-Ranked workflow would require 2,990 unnecessary rounds if the model stabilized after 10 rounds of Top-Ranked Active Learning.

## III. EXPERIMENTAL PROCEDURE

Aside from Cormack and Grossman's work [3][4][5], few published studies exist that research Active Learning training document selection strategies in the legal domain. The purpose of the work reported in this paper is to empirically compare Random sampling document selection with two Active Learning document selection strategies, Top-Ranked and Uncertain, in the context of document review.

We conducted experiments on five data sets from confidential, non-public, real legal matters across various industries such as social media, communications, education, and security. We chose matters with data sets that ranged from 100,000 to 500,000 documents in order to execute our experiments within a reasonable time period. The richness, or positive class rate, of the five data sets ranged from approximately four percent to 40 percent. Attorneys reviewed all documents in the five data sets over the course of the legal matter and their review-coding labels provided the ability to evaluate the performance of the models at any time. Tables 1A, 1B, and 1C provide detail about the data sets including descriptions, their size, and attorney coding statistics. The predictive modeling objective for Data Sets A, B, and C was to identify Responsive documents. The modeling objective for Data Sets D and E was to identify Privileged documents. These data sets represent five real-world scenarios, occurring between 2010 and 2016, and provide a feature rich environment with which to perform our experiments. They are compelling since they are diverse in content, have labels for Responsiveness and also Privilege reviews, and have not been used in previous legal domain research like the TREC 2009 Legal Track study [12] and by Cormack and Grossman [3][4]. Many legal domain text analytics research studies use the publicly available Enron email data set produced to the Federal Energy Regulatory Commission (e.g., [3][4][12]). This data set is more than 15 years old and may not well represent current legal matter email communications. Finally, the data sets range of richness rates allowed us to study the richness rate's impact on the effectiveness of the learning strategies.

Table 1A: Data Set Description

| |
|---|
| A – Emails with attachments, and individual Microsoft Office and PDF documents relating to an intellectual property and trade secret matter. |
| B – Emails with attachments relating to a U.S. enforcement agency investigation. |
| C – Emails with attachments, and individual Microsoft Office and PDF documents relating to a U.S. enforcement agency investigation. |
| D – Emails with attachments relating to a U.S. regulatory agency investigation. |
| E – Emails with attachments relating to a U.S. regulatory agency investigation. |

Table 1B: Responsive Data Set Statistics

| Data Set | Total Documents | Responsive Documents | Not Responsive Documents | Richness |
|---|---|---|---|---|
| A | 170,301 | 53,546 | 116,755 | 31.4% |
| B | 502,050 | 201,906 | 300,144 | 40.2% |
| C | 246,058 | 10,214 | 235,844 | 4.2% |

Table 1C: Privileged Data Set Statistics

| Data Set | Total Documents | Privileged Documents | Not Privileged Documents | Richness |
|---|---|---|---|---|
| D | 360,531 | 46,756 | 313,775 | 13.0% |
| E | 397,289 | 14,326 | 382,963 | 3.6% |

Our training document selection strategies were defined as:

- **Top-Ranked** documents were the *N* documents with the highest scores assigned by the model.
- **Uncertain** documents were the *N* documents with scores that were closest to 0.5 from a range of scores between 0.0 and 1.0.
- **Random** *N* random documents were sampled regardless of score.

Our experiments evaluated the performance of each strategy using the following metrics: recall, the percentage of documents requiring review, and Optimum Performance. Optimum Performance is a metric created by this group of collaborators to confirm the Active Learning round that a document selection strategy is most effective at reducing the volume of documents for manual review. At the Optimum Performance round, the fewest number of documents (including the documents reviewed for training) require review to achieve a specified recall. Within the legal domain, predictive coding typically aims to minimize the documents that require review in order to achieve a specified recall. In document productions, responsive recall rates generally range between 70 percent and 90 percent depending on the matter and the data set, with recall rates of 75 percent to 80 percent often accepted by courts and litigants. Accordingly, attorneys and their clients seek to review only enough documents in the document population to meet the specified level recall – making the percentage of documents requiring review and Optimum Performance important performance metrics. These metrics directly relate to the cost effectiveness of a document selection strategy and their practical implementation.

The machine learning algorithm we used to generate models was Logistic Regression. One of our prior studies demonstrated that predictive models generated with Logistic Regression perform very well on legal matter documents [13]. Other parameters we used for modeling were, bag of words with 1-gram and normalized frequency, and 20,000 tokens were used as features.

While legal data is not always huge, the active learning process requires Big Computation for both training and scoring. The active learning process must repeat hundreds of times for each data set to finalize the review process. Effectively, the active learning process requires hundreds of trained models and towards the end of the process those models are trained hundreds of thousands of training documents. Models created in those later rounds can take many hours to generate and we use multiple machines to distribute the processing.

On average and for each dataset, our active learning rounds required scoring more than 100,000 documents and completing the entire process required scoring more than 10 million documents. To score documents efficiently, we used Lucene to index all the scoring documents in every round of the active learning process. We did this so we did not have to read the scoring documents into memory for each round. Additionally, instead of reading and scoring document by document, the scoring algorithm scores all documents together, one term after another, using the Lucene inverted index lists. This implementation dramatically reduced the scoring time.

We conducted two types of experiments. With Type One, we randomly selected 10 percent of the documents in each data set and generated a validation set. Active Learning was then executed on the remaining 90 percent of the data set. Finally, the predictive model was evaluated using the validation set after each round of the Active Learning process. Type One experiments were designed to empirically compare the performance of the models generated using different strategies. Table 2 outlines the process of the Type One experiments.

Table 2: Type One Experimental Procedure

| |
|---|
| - Randomly select 10% of the data set for validation. |
| - Using the remaining 90% of the data set, randomly select 1,000 training documents. |
| - Train a model. |
| - Evaluate the model using the validation set. |
| - Score all remaining documents excluding the validation and training documents using the model. |
| - Add 1,000 documents using one of the three selection strategies to the training set from the remaining 90% of the data set not previously used for training. |
| - Repeat Step 3 through 6 until there are no more documents left in the 90% of the data set. |

The second type of experiments, Type Two, evaluates the models generated using the three strategies after each round using all documents including both training documents and all remaining documents. The performance of models generated in the Type Two experiments cannot be compared using only the remaining documents because the population of remaining documents differs across strategies – each strategy selects a different set of training documents round to round.

Different from most other studies, performance metrics in our Type Two experiments were computed after each Active Learning round until 100 percent recall was achieved. After each round, the data set was partitioned to two sets of documents. The first set contained the previously reviewed documents selected for training. The second set contained the remaining documents. The training documents increases with the rounds of Active Learning. The model developed after each round was applied to score all the remaining documents and a cut-off score is then used to categorize the remaining documents as Responsive or Privileged, namely the documents with probability scores greater than or equal to the cut-off score. The cut-off score was determined according to the recall requirement. The documents with scores above or equal to the cut-off score are the documents that should be reviewed by attorneys in order to achieve a specified recall. The experiments repeated until the specified recall was achieved within the documents selected for training by the training selection strategy. Table 3 outlines the process of the Type Two experiments.

Table 3: Type Two Experimental Procedure

| |
|---|
| 1. Randomly select 1,000 training documents. |
| 2. Train a model. |
| 3. Score all the remaining documents using the model. |
| 4. Evaluate the model using both training documents and the remaining documents. |
| 5. Add 1,000 documents selected using one of the three selection strategies to the training set. |
| 6. Repeat Step 2 through 6 until 100 percent recall is achieved in the training set. |

The performance of a model in Type Two experiments was computed using both documents reviewed (training documents) and documents to be reviewed (documents categorized as Responsive or Privileged by the model). To demonstrate an example of a Type Two experiment performance calculation using 10 active learning rounds, assume that:

- The total number of documents in the data set was 300,000.
- The total number of Responsive documents in the data set was 50,000.
- We wanted to achieve 75 percent recall.
- We conducted 10 rounds of Active Learning using one training document selection strategy.
- A total of 10,000 documents were reviewed for training in the 10 rounds (1,000 documents per round) of Active Learning and 7,000 of the training documents were deemed Responsive.

We applied the predictive model generated after 10 rounds of Active Learning to score the remaining documents in the data set, excluding the 10,000 training documents. We then determined the cut-off score needed to achieve 75 percent recall. In this example, assume there are 60,000 documents above the cut-off score. Our calculations were as follows:

- 37,500 Responsive documents have to be identified in order to achieve 75% recall
  - 0.75 * 50,000 = 37,500 Responsive documents
- However, 7,000 Responsive documents were identified in the 10 rounds of Active Learning, therefore 30,500 Responsive documents must have scores greater than or equal to the cut-off score
  - 37,500 – 7,000 = 30,500 Responsive documents
- The percentage of documents requiring review is then computed as the number of documents above the cut-off score (60,000) plus 10,000 (training documents reviewed) divided by the total number of documents in the data set
  - (60,000 + 10,000) / 300,000 = 23.3% of the documents require review

IV. EXPERIMENT RESULTS

In this section, we report and discuss our experimental results using Top-Ranked, Uncertain, and Random strategies to select training documents. We report the percentage of documents requiring review for each training document selection strategy at various recall rates. The process, described in III. EXPERIMENTAL PROCEDURE, was created to confirm which strategy would achieve the target recall rates while reviewing the fewest number of documents.

A. *Type One Experiments*

In this section, we report experimental results from the Type One experiments. In these experiments, models were evaluated using the validation set – a 10 percent random sample from each data set.

Figure 1 details the percentage of documents requiring review for each round of the Active Learning process at 75 percent and 90 percent recalls. Our results clearly demonstrate that the Top-Ranked strategy does not perform as well as the Uncertain and Random strategies in most cases. In Data Sets A, B, and D, Uncertain and Random significantly outperform Top-Ranked at both 75 and 90 percent recalls. Top-Ranked converges slowly and it needs many more rounds of Active Learning than Uncertain and Random in order to achieve the optimum performance. This makes perfect sense because, as described in II. ACTIVE LEARNING, documents selected using the Top-Ranked strategy have similar content patterns to those in each round's training set, which may provide very little new information that the algorithm can use to generate improved models. In contrast, the documents selected using the Uncertain and Random strategies are typically dissimilar to the documents in the previous round's training set and likely allow the algorithm to detect new content patterns. For highly imbalanced Data Sets C and E, Top-Ranked and Uncertain perform similarly. Because the number of positive documents is small and not many documents received large scores, Top-Ranked and Uncertain selected similar training documents. Especially after a few rounds of Active Learning, many of the positive documents were already selected as training documents. Random performs better than Top-Ranked and Uncertain, because Random selected a wide range of training documents. Surprisingly, Random performs almost as well as or better than Uncertain in all data sets. Additionally, we observed that Top-Ranked outperforms Uncertain and Random strategies when low recall is acceptable; however, when high recall is the goal, Uncertain and Random perform better than Top-Ranked [14].

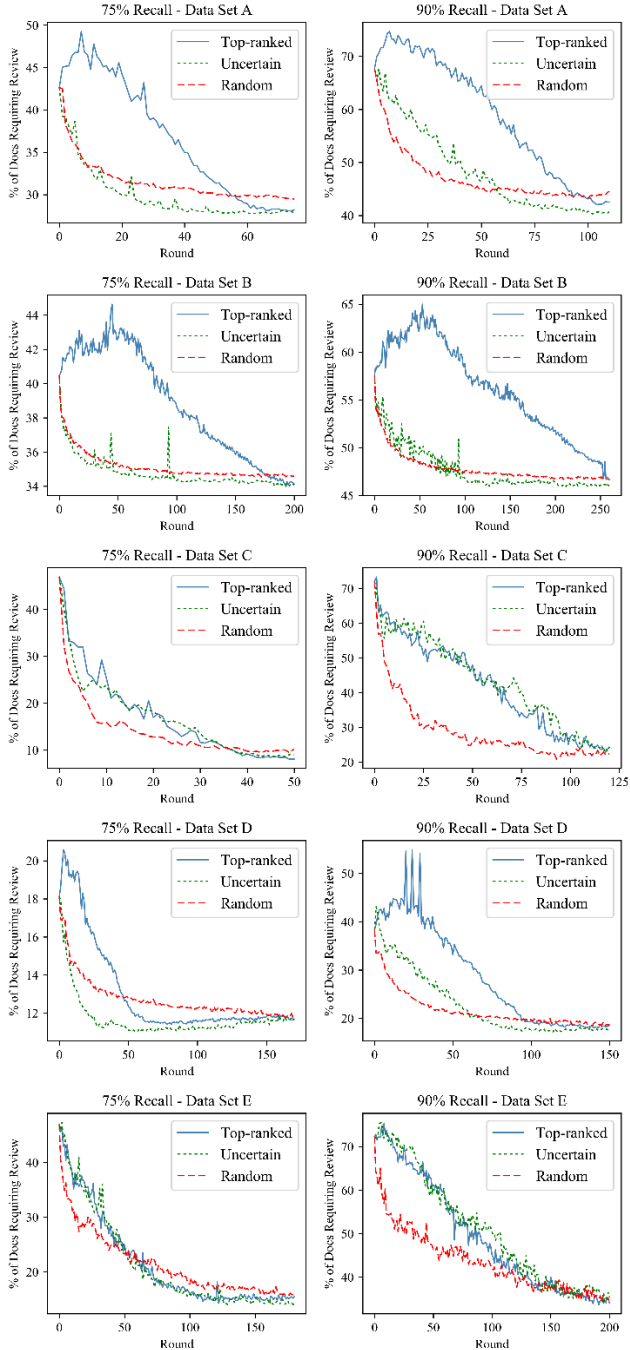

Figure 1. Type One Experiment Performance

## B. Type Two Experiments

### 1. Recall Achieved During Training Set Review

The Figure 2 provides round by round details for the recall of the documents within the training set for each training document selection strategy. The recalls for Top-Ranked are significantly higher than the recalls for the other two strategies in Data Sets A, B, and D. Uncertain obtains higher recalls than Random. These results are intuitive. Clearly, top-ranked (high-scoring) documents include more Responsive or Privileged documents than documents with scores around 0.5 and randomly selected documents. For Data Sets C and E, the results for the Top-Ranked and Uncertain strategies are similar. The richness rates for these sets were low, approximately four percent, and very few documents received high scores. Therefore, most of the Uncertain documents were also Top-Ranked, with scores close to 0.5, resulting in similarly selected training documents between the two strategies. The recalls of Uncertain are better than Random, with Random increasing linearly as new training documents are added each round.

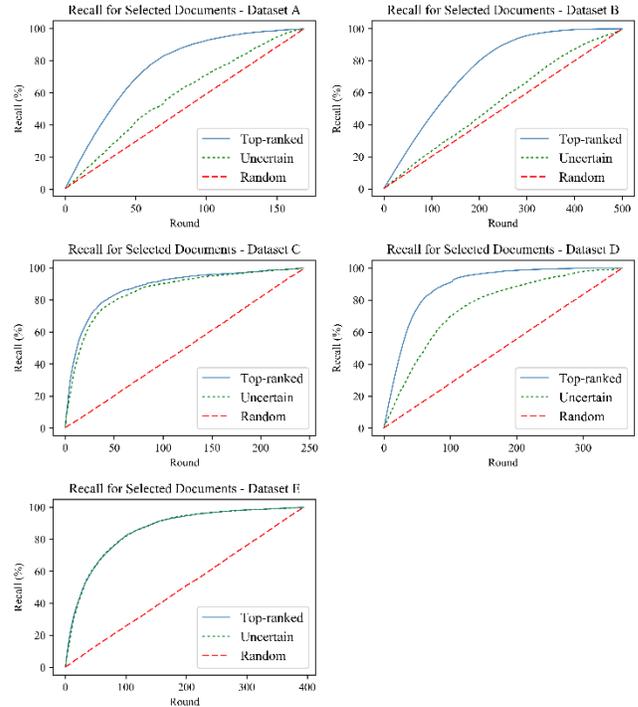

Figure 2. Round by Round Recall Achieved During Training

### 2. Percentage of Documents Requiring Review

Figures 3 details the percentage of documents requiring review for each round of the Active Learning process for two recall rates: 75 percent and 90 percent. We compare the performance of the three training document selection strategies and also the point at which the Active Learning process achieved the Optimum Performance. Described in section 4.1, Optimum Performance is the active learning round that a strategy is most effective at reducing the volume of documents for manual review.

The performance patterns were similar in Data Sets A, B, and D. As illustrated in the figures, Uncertain performed the best overall, reaching Optimum Performance much earlier than the Top-Ranked strategy for all data sets. Its percentages of documents requiring review are slightly lower than the Top-Ranked strategy other than Data Set B at 75 percent recall. This is directly related to the superior performance of Uncertain strategy's predictive models demonstrated in the Type One experiments. The advantages of Uncertain over Top-Ranked are more obvious at 90 percent recall than 75 percent recall. Random can quickly achieve the Optimum Performance across the data sets but its overall Optimum Performance is still inferior to Uncertain and Top-Ranked.

For Data Sets C and E, as observed in the Type One experiments, the results for the Top-Ranked and Uncertain strategies were similar likely due to their low richness rates and lack of high scoring documents.

After reaching Optimum Performance (the lowest point on the curves), the percentage of documents requiring review increased with the use of the Uncertain and Random strategies. This performance behavior was in contrast to the behavior of Top-Ranked and occurred for two reasons:

- The models developed in the first few rounds were accurate enough to identify 75 percent or 90 percent of the Responsive or Privileged documents when using model predictions.
- There may not be as many Responsive or Privileged documents in the documents selected for training using the Random or Uncertain strategies when compared to Top-Ranked. As active learning rounds increase, the number of Responsive or Privileged documents decreases.

This performance behavior does not imply the models cannot be improved or that they have stabilized. It simply means that it may not be worth trying to improve the model by continuing the Active Learning process. On the other hand, the Top-Ranked performance continues to improve, but not by much.

Like the Type One experiments, in early rounds the Random strategy performed better than the other strategies for Data Sets C and E, demonstrating that the strategy likely performs well on low richness data sets. This is because the strategy selects random documents across the entire range of scores and on Data Sets C and E specifically, the strategy selected more informative training documents early on making its models more effective more quickly.

*3. Optimum Performance*

In our Active Learning experiments, the Optimum Performance is the point in the process that a strategy is most effective at reducing manual document review volumes. At this point, the fewest number of documents require review. We tracked the round that optimum performance occurred for each training document selection strategy and used this to compare the results of the selection strategies. Training Set Recall was another performance metric we used. The Training Set Recall is the recall calculated using only the documents selected and reviewed as training documents at the Optimum Performance point.

Tables 4 and 5 detail the Optimum Performance statistics for each training document selection strategy at 75 percent and 90 percent recall, respectively. Our initial observation was that the Top-Ranked strategy requires the Active Learning process to nearly complete in order to reach Optimum Performance. For example, to achieve 90 percent recall on Data Set B, the Top-Ranked Active Learning process continued for 248 rounds, while the Uncertain strategy achieved 90 percent recall in 63 rounds. In a real legal document review, the Top-Ranked strategy would have required 185 unnecessary rounds of modeling when compared to the Uncertain strategy.

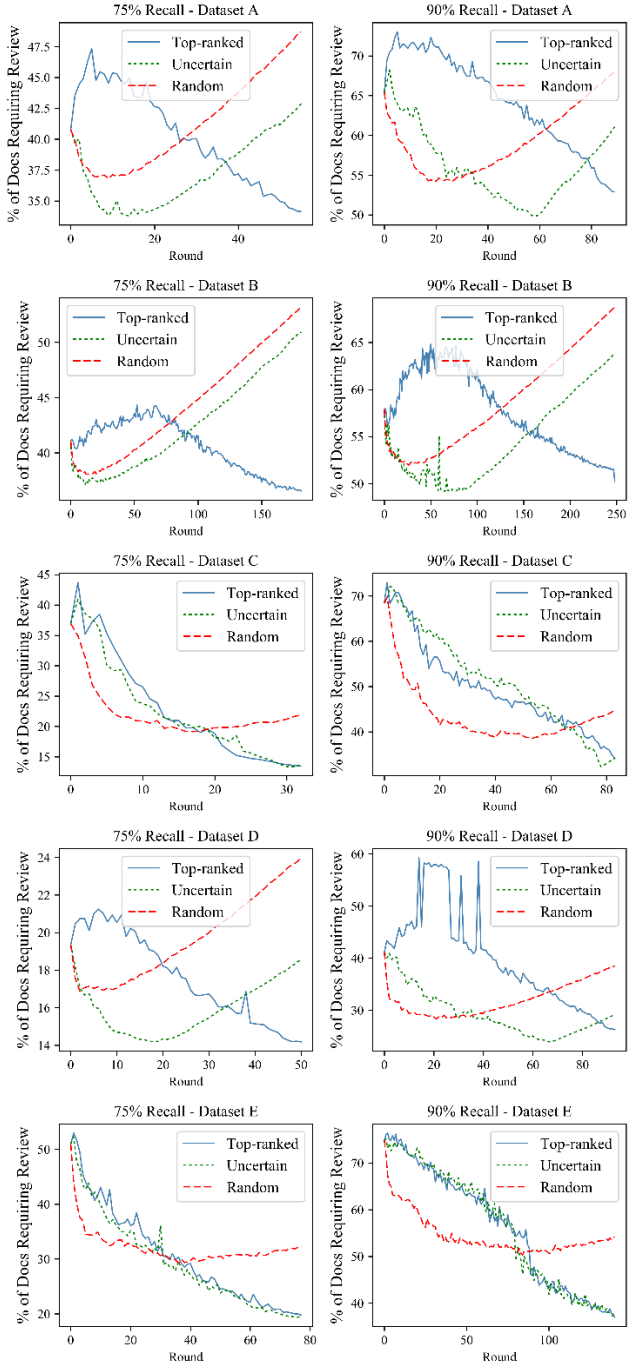

Figure 3. Round by Round Percentage of Documents Requiring Review

best performing models have different characteristics at 75 percent recall and 90 percent recall, (ii) the selected training documents are more accurate than the model predictions (documents yet to be selected for training), or, (iii) both possibilities are true. More analysis and experimentation is needed to understand the observation. This observation has a practical implication though: more rounds of Active Learning must be performed to achieve higher recall rates.

The Uncertain strategy achieved Optimum Performance earlier than the Top-Ranked strategy. The models trained with documents selected using the Uncertain strategy are much more effective when compared to the models trained with documents selected using the Top-Ranked strategy. As described in II. ACTIVE LEARNING, documents selected using the Top-Ranked strategy may provide very little new information that the learning algorithm can use to generate improved models. In contrast, the Uncertain strategy identifies the most informative training documents to enhance the models' effectiveness. Models built using the Random strategy performed very well, too.

Illustrated in Table 5, the Top-Ranked strategy's Training Set Recall contained nearly 90 percent recall for all data sets, meaning that almost all of the Responsive or Privileged documents were identified in the Active Learning process, thereby removing the option to use model predictions. It should be noted that the Uncertain and Random strategies' Training Set Recalls contained 15.5 percent and 5.4 percent, respectively, with the remaining recall coming from the documents predicted by the model.

Table 4: Optimum Performance Statistics for 75% Recall

| Data Set | Selection Strategy | Optimum Performance Round | Training Set Recall | % of Documents Requiring Review |
|---|---|---|---|---|
| A | Top-Ranked | 55 | 73.5 | 34.1 |
| A | Uncertain | 14 | 12.3 | 33.8 |
| A | Random | 9 | 6 | 36.8 |
| B | Top-Ranked | 181 | 74.5 | 36.5 |
| B | Uncertain | 11 | 2.9 | 37.1 |
| B | Random | 13 | 2.8 | 38.1 |
| C | Top-Ranked | 32 | 74.8 | 13.5 |
| C | Uncertain | 30 | 68.6 | 13.4 |
| C | Random | 17 | 6.9 | 19.2 |
| D | Top-Ranked | 50 | 74.9 | 14.2 |
| D | Uncertain | 18 | 16.7 | 14.2 |
| D | Random | 2 | 0.8 | 16.9 |
| E | Top-Ranked | 77 | 74.7 | 19.8 |
| E | Uncertain | 76 | 74.7 | 19.4 |
| E | Random | 36 | 9.6 | 29.3 |

Interestingly, Optimum Performance was achieved in earlier rounds for 75 percent recall than for 90 percent recall in all five data sets. One might have assumed that Optimum Performance would be achieved in a similar round for the Uncertain and Random strategies when comparing the two recalls because the models should stabilize at the same time. Initial reactions to this result lead us to think that, (i) the

Table 5: Optimum Performance Statistics for 90% Recall

| Data Set | Selection Strategy | Optimum Performance Round | Training Set Recall | % of Documents Requiring Review |
|---|---|---|---|---|
| A | Top-Ranked | 89 | 90 | 52.9 |
| A | Uncertain | 59 | 48.2 | 49.9 |
| A | Random | 26 | 15.9 | 54.2 |
| B | Top-Ranked | 248 | 89.6 | 50.2 |
| B | Uncertain | 63 | 15.5 | 49.2 |
| B | Random | 26 | 5.4 | 52 |
| C | Top-Ranked | 83 | 89.8 | 34.2 |
| C | Uncertain | 78 | 87.3 | 32.3 |
| C | Random | 53 | 21.5 | 38.4 |
| D | Top-Ranked | 93 | 89.9 | 26.3 |
| D | Uncertain | 67 | 53.3 | 23.9 |
| D | Random | 21 | 5.9 | 28.3 |
| E | Top-Ranked | 140 | 89 | 37 |
| E | Uncertain | 146 | 89.8 | 37.1 |
| E | Random | 84 | 21.9 | 50.3 |

## V. CONCLUSIONS

Active Learning has drawn the attention of the legal community because of its potential to make the predictive coding process even more effective and efficient. Legal practitioners with text analytics expertise increasingly use Active Learning in lieu of Passive Learning and more traditional forms of training. The purpose of our study and this paper is to explore and highlight the ways Active Learning can be most effective when used in the legal domain.

Our experiments empirically compared a random sampling training document selection strategy with two Active Learning strategies, Top-Ranked and Uncertain, in the context of document review. We compared the performance of predictive models generated using the three training document selection strategies. Unlike many other studies in this area, our study focused on reducing the volume of documents requiring manual review. Such review accounts for the vast majority of costs in litigation. Training document selection strategies that reduce the number of documents needing manual review result in significant cost savings to the attorneys' clients. Accordingly, we developed a new metric – Optimum Performance – to measure the point in the Active Learning process that a training document selection strategy requires manual review of the smallest percentage of documents. Our study provided many observations with three highlights below:

- The Top-Ranked strategy requires the most Active Learning rounds to achieve its Optimum Performance at 75 and 90 percent recalls.
- When compared to Top-Ranked, the Uncertain strategy achieved its Optimum Performance with fewer rounds across all data sets and each recall rate, except for Data Set E at 90 percent recall. This indicates that using Active Learning throughout the entire document review is less efficient than stopping the process once an acceptable model performance is achieved.
- Random sampling is an effective learning method when considering its Optimum Performance, compared to that of Top-Ranked. Given the legal community's bias toward Active Learning, our results were very interesting and debunk the myth that Active Learning with Top-Ranked are superior machine learning techniques. The great performance of the Random strategy in low richness data sets suggests a future work idea, a hybrid strategy that combines Top-Ranked, Uncertain, and Random.

In sum, our study yields insights about the use of Active Learning in predictive coding for document review. While the selection of training documents for predictive coding has historically been a guessing game for legal practitioners, the results of our research shed light on the most effective training document selection strategies. When using the Passive Learning approach, there is no longer a need to identify training documents using educated guesses. For example, legal teams should consider initially generating their models using random samples for low richness matters and not Top-Ranked – although, this is contrary to current popular opinion in the legal community.

Our observations call for the creation of a more sophisticated predictive coding methodology, one that combines Active Learning to train the model until it reaches Optimum Performance and then relies on model predictions to reach the desired target recall. This team of collaborators plans to outline this new methodology in the future, but for now, legal practitioners can use these observations to evaluate their current Active Learning and predictive coding processes. By implementing the insights offered in this paper, legal practitioners can improve their predictive coding outcomes significantly.